\documentclass[reprint,longbibliography,amsmath,amssymb,amsfonts,prd,superscriptaddress]{revtex4-1}
\usepackage{graphicx}
\usepackage{dcolumn}
\usepackage{bm}
\usepackage[colorlinks=true, citecolor=blue, linkcolor=blue, urlcolor=blue]{hyperref}
\usepackage{xcolor}
\usepackage{amsmath, amssymb, amsfonts}
\usepackage{mathtools}
\usepackage{subfigure}
\usepackage{mathrsfs}
\usepackage{bbold}
\usepackage{sidecap}
\usepackage{verbatim}

\definecolor{purple1}{rgb}{128,0,128}

\newcommand{\nn}{\nonumber\\}
\usepackage{bigints}
\newcommand{\bea}{\begin{eqnarray}}
\newcommand{\ea}{\end{eqnarray}}
\newcommand{\ord}{{\cal O}}

\definecolor{darkpastelgreen}{rgb}{0.01, 0.75, 0.24}
\begin{document}
\title{Analogue gravitational field from nonlinear fluid dynamics} 
\author{Satadal Datta} 
\author{Uwe R. Fischer}
\affiliation{Seoul National University, Department of Physics and Astronomy, Center for Theoretical Physics, Seoul 08826, Korea}
\date{\today}

\begin{abstract}
The dynamics of sound in a fluid is intrinsically nonlinear. 
We derive the consequences of this fact for the analogue gravitational field experienced
by sound waves, by first describing generally how the nonlinearity of the equation for phase fluctuations back-reacts onto the definition of the background providing the effective space-time 
metric. 
Subsequently, we use the the analytical tool of Riemann invariants 
in one-dimensional motion to derive source terms of the effective gravitational field 
stemming from nonlinearity. Finally, we show that the consequences of 
nonlinearity we derive can be observed with Bose-Einstein condensates in the ultracold gas laboratory.
\end{abstract}

\maketitle
\section{Introduction}



The basic assumption underlying the simulation of curved space-times, coined 
analogue gravity \cite{BLV}, is the separation 
of the underlying 
classical or quantum field into a perturbation field and a background field. 
Such a separation is conventionally possible when one has a control 
parameter (such as the number of particles in mean-field approaches to condensed 
matter systems),  which separates the field into a large background and its perturbations.
In addition, one needs to separate the length and frequency scales 
of background and perturbation to have them well defined and separable.
Under rather general conditions for the action of the system under consideration, then 
a wave equation results which is identical to that 
of a scalar (in the simplest case), minimally coupled to gravity
\cite{BLV_normal}. 
This standard paradigm, originally due to Unruh \cite{unruh} (see also for an early precursor \cite{Trautman}) has yielded analogues of classical and quantum field propagation in flat and curved space-time for a multitude of physical contexts,
e.g., analogues  of Lorentzian signature space-times and the associated kinematical effects of
quantum fields on these,  experimentally as well as theoretically. 
A non-exhaustive list of examples comprises black holes via (shallow water) gravity waves \cite{RalfBill,Weinfurtner,Euve,EuveII}, black holes in fluids of light \cite{Marino,Nguyen}, numerous studies on Hawking radiation, e.g., \cite{Carusotto_2008,Macher,Gerace,Steinhauer16,Munoz}, the inflationary Universe  
and Hubble dynamics \cite{Schuetzhold,PhysRevLett.118.130404,Eckel,Eckel2,Banik}, the pair-production of cosmological quasiparticles \cite{BLV2003PRA,CPP,steinhauer2021analogue} and their associated degree of entanglement \cite{Busch,Robertson,Tian}, the Unruh and Gibbons-Hawking effects  as manifestations of the observer dependent content 
of quantum fields in flat and curved (de Sitter) space-time \cite{Fedichev,Reznik,Celi,Gooding}, 
the quantum back-reaction on a classical background \cite{RalfPRD}, and
to probe analogue trans-Planckian effects on low-energy phenomena \cite{PhysRevLett.118.130404,TianDu}. 
Furthermore, this standard paradigm has been harnessed to investigate the black hole lasing phenomenon for black-white hole configurations \cite{Ted,Finazzi}, black hole  
superradiance \cite{Basak,Torres,Prain} and  quasinormal black hole 
modes \cite{Richartz}, as well as analogues of gravitational waves  \cite{hartley2018analogue,PhysRevD.98.064049}, and to address aspects of the black hole information paradox \cite{Andrea}. 

The underlying nonrelativistic medium in the laboratory, in its continuum description, is usually however intrinsically nonlinear.
For instance, one encounters, in the case of a fluid, in the course of time unavoidably the 
fluid-dynamical nonlinearity will enter the dynamics of the fluid velocity and density, 
and the basic linearization premise on which conventional analogue gravity is based will break down.
Even when initially a linear description applies, eventually the nonlinear dynamics of the 
fundamental variables becomes manifest, and finally a shock wave singularity will develop. 

The standard paradigm of quantum field theory, leading to the definition of 
(quasi-)particles, is in fact precisely this 
linearization procedure on top of an essentially inert background.
It underlies the majority of derivations of phenomena which are described by 
quantum fields propagating in {\it fixed} curved space-time \cite{Birrell}.
However, for concreteness, in the arena of the nonlinear dynamics of fluids, there are 
only two variables, density and velocity of the fluid, and the separation into background 
and perturbations, when one goes beyond simply linearizing the equations governing
the perturbations, needs to be readdressed. 

Our aim in this paper is to address 
how the intrinsic nonlinearity of perfect fluid 
dynamics affects the concept of analogue gravity and the definition of the space-time metric
which is attached to the background field.
We show, in particular, that the space-time metric which affords a suitable description of
the effective analogue gravitational field which furnishes the wave equation in curved space-time 
changes when one has to take into account that the background solution derives from the 
solution of the full nonlinear fluid-dynamical equations. 
As a further consequence of nonlinearity, we then demonstrate dynamical aspects of metric perturbations above a background metric, consisting in the emergence of source terms
in the wave equation for the metric perturbations.

To treat the problem of fluid-dynamical nonlinearity in a tractable manner, we then 
consider the 1+1D of Riemann wave equation and the theory of Riemann invariants, 
which are furnishing an analytical description of the emergence of shock waves.
Using the Riemann approach, we obtain source terms which are constituting sources 
of the propagating gravitational perturbation field (such as gravitational waves), which sources
themselves depend nonlinearly on the metric perturbations.  
We are thus supplying a concrete experimentally testable setup, in an analogue gravity setup,  for the emergence of a curved space-time metric from a Minkowski metric due to the nonlinearity of the underlying (scalar) field theory \cite{Novello_2011}.  

We finally provide concrete estimates for the experimental manifestations of such a nonlinear
analogue gravity in Bose-Einstein condensates (BECs).
In particular, we show that the time periods $t_{\rm shock}$ for shock waves to emerge,
 for current BEC setups, also and in particular those which study analogue gravity phenomena,
 are much less than the lifetime of typical experimental runs. Hence we argue that, when studying  analogue gravity, the nonlinearity of fluid dynamics in the perfect fluid BEC must generally 
 be taken into account.

\section{Action Principle for Nonlinear Fluid Dynamics}




We assume in the following that we are in the nondispersive limit of the fluid dynamics, 
which for a BEC in particular implies the neglect of the quantum pressure  term involving
density gradients (the so-called Thomas-Fermi limit in BECs).  
The action for an inviscid irrotational barotropic fluid 
then is \cite{stone},  
\begin{equation}
S= -\int d^4x\left\{ \rho \dot \Phi +\frac{1}{2} \rho (\nabla\Phi)^2
+ u(\rho)+V_{\rm ext}\rho\right\}. \label{action}
\end{equation}
Here, $\rho({\bm x},t), \Phi({\bm x},t)$, $V_{\rm ext}({\bm x},t)$ are fluid density, velocity potential and a scalar potential corresponding to external conservative force, respectively; $u(\rho)$ is internal energy density.
The Inviscid irrotational fluid equations of motion are derivable by varying the above action:
\begin{eqnarray} 
& \partial_t\rho+\rho\nabla\cdot {\bm v}+{\bm v}\cdot\nabla\rho =0, \label{continuity}\\
&\rho\partial_t {\bm v}+\rho{\bm v}\cdot \nabla{\bm v}+c_s^2{\nabla \rho}+\rho\nabla V_{\rm ext}=0,  \label{euler}
\end{eqnarray}
where fluid velocity ${\bm v}({\bm x},t)=\nabla{\Phi}$, fluid pressure $p({\bm x},t)=-\left(\rho \dot \Phi +\frac 12 \rho (\nabla\Phi)^2
+ u(\rho)\right)$, and $c_s^2=c_s^2 (\rho)=\frac{dp}{d\rho}$, where $c_s$ is  the sound speed.
The fluid equations Eq.~\eqref{continuity}-\eqref{euler} represent a system of first order quasilinear partial differential equations, a particular type of nonlinear partial differential equation  \footnote{In a quasilinear partial differential equation,, the highest order derivatives of dependent variables occur linearly, with their coefficients functions of only lower order derivatives, whereas any term with lower order derivatives can occur nonlinearly.}.
A well posed boundary value problem gives a unique solution in the domain of ${\bm x}-t$. 
If one considers, in particular, two solutions of $\rho$ and ${\bm v}$ originating from two boundary value problems, one may select one of them as the background. 
Therefore, in general the definition of a background is arbitrary. Conventionally, 
one selects a solution which varies slowly with ${\bm x}$, and $t$ as a background or ``mean" flow. 
Any solution of flow can be decomposed into a mean flow plus a perturbation terms in density and velocity. For a given solution chosen as the background flow, a variation in the boundary value problem produces perturbation terms ($\delta\rho$ and $\delta{\bm v}$). 
We then expand the action in Eq.~\eqref{action} as follows, 
\begin{equation}
S=S_0+\delta S.
\end{equation} 
Here, $S_0$ is the action corresponding to the background flow. The term linear in perturbations, vanishes because the background by definition satisfies itself the fluid-dynamical equations. Hence, $\delta S$ consists of a term quadratic in perturbations and higher order terms.
Thus we have  
\begin{eqnarray}
& \delta S=-\int d^4x\left\{ \delta\rho \delta\dot \Phi +\delta\rho{\bm v}_0\cdot\nabla\delta\Phi+\frac{1}{2}\rho_0(\nabla\delta\Phi)^2\right.\nonumber\\ 
&+\left.\frac{1}{2} \delta\rho (\nabla\delta\Phi)^2
+ \delta\psi\right\}. \label{action2}
\end{eqnarray}
We denote the background quantities with suffix $0$, and $\delta\psi$ is the series of terms in $\delta u$ starting from terms of order $\delta\rho^2$ and higher orders.  
\begin{eqnarray}
& u(\rho)=u(\rho_0+\delta\rho)=u(\rho_0)+\delta u\nonumber\\
& =u(\rho_0)+\frac{du}{d\rho}|_{\rho=\rho_0}\delta\rho+\frac 12 \frac{d^2u}{d\rho^2}|_{\rho=\rho_0}\delta\rho^2+..\nn 
&=u(\rho_0)+\frac{du}{d\rho}|_{\rho=\rho_0}\delta\rho+\delta\psi.
\end{eqnarray}
The specific enthalpy, $h(\rho)=\frac{du}{d\rho}\Rightarrow\delta h= \frac{\partial\delta\psi}{\partial\delta\rho}$.
$\delta\rho$ can be found from the equation of motion of $\delta\rho$ in the action, $\delta S$ of Eq.~\eqref{action2}:
\begin{equation}\label{dh}
\delta\dot{\Phi}+{\bm v}_0\cdot\nabla\delta\Phi+\frac{1}{2}  (\nabla\delta\Phi)^2+\delta h=0,
\end{equation}
along with $h=\int\frac{dp}{\rho}$ and $\delta h=h(\rho)-h(\rho_0)$.
Therefore, the inverse function of $h$ (assuming its existence) reads 
\begin{multline}\label{h}
h^{-1}\left(\frac{h(\rho)}{h(\rho_0)}\right)\\ 
 = h^{-1}\left(1-\frac{1}{h(\rho_0)}\left(\delta\dot{\Phi}+{\bm v}_0\cdot\nabla\delta\Phi+\frac{1}{2}  (\nabla\delta\Phi)^2\right)\right).
\end{multline}
 We consider the simplest possible case, such that $\frac{h(\rho)}{h(\rho_0)}=h({\rho}/\rho_0)$. Therefore $h(\rho)\propto\rho^q$, $q$ is some real number. This is the case for an ideal gas with polytropic equation of state: $p=K\rho^\gamma$, where $K$ is a constant, 
 and $\gamma$ is the ratio of the specific heat capacities; then $q=\gamma -1$.
We may then functionally express the density variations as 
\begin{equation}\label{10}
\delta\rho= 
\rho_0(\mathscr{F}^{\frac{1}{\gamma -1}}-1),
\end{equation}
where the functional $\mathscr{F}$ is given by 
\begin{equation}\label{11}
\mathscr{F}\left(\delta\dot \Phi, \nabla{\delta\Phi}\right)
=1-\frac{\gamma -1}{c_{s0}^2}\left(\delta\dot{\Phi}+{\bm v}_0\cdot\nabla\delta\Phi+\frac{1}{2}  (\nabla\delta\Phi)^2\right).
\end{equation}
Thus $\delta\rho=\delta\rho \left(\delta\dot \Phi, \nabla{\delta\Phi}, {\bm x}, t \right)$. The 
perturbation Lagrangian density $\delta\mathscr{L}$ corresponding to the action $\delta S$ is:
\begin{eqnarray}\label{LD}
&\delta\mathscr{L}\left(\delta\dot \Phi, \nabla{\delta\Phi}, {\bm x}, t \right))
\nonumber\\
&=-\rho_0(\mathscr{F}^{\frac{1}{\gamma -1}}-1)
\left(\delta\dot{\Phi}+{\bm v}_0\cdot\nabla\delta
\Phi+\frac{1}{2}  (\nabla\delta\Phi)^2\right)\nonumber\\
&+\frac{1}{2}\rho_0(\nabla\delta\Phi)^2+\delta\psi.
\end{eqnarray} 
Therefore, using the expression \eqref{11},
\begin{eqnarray}\label{LD2}
& \delta\mathscr{L}\left(\delta\dot \Phi, \nabla{\delta\Phi}, {\bm x}, t \right)=-\frac{1}{2}f^{\mu\nu}_0\partial_\mu\delta\Phi\partial_\nu\delta\Phi+\delta\mathscr{L}_I,
\end{eqnarray}
where we define the coefficient matrix $(\mu,\nu=t,x,y,z)$ 
\begin{equation}
f^{\mu\nu}_0\coloneqq\frac{\rho_0}{c_{s0}^{2}}\begin{bmatrix}
-1 & \vdots & -{\bf v}_0 \\
\cdots&\cdots&\cdots\cdots \\
-{\bf v}_0^T&\vdots & c_{s0}^{2}\mathbb{I}-{\bf v}_0\otimes{\bf v}_0 .
\end{bmatrix}. \smallskip 
\end{equation}
Note that the self-interaction part $\delta\mathscr{L}_I$ of Eq.~\eqref{LD2} 
is an infinite series in $\delta\dot \Phi$ and $\nabla{\delta\Phi}$, 
except for the particular (BEC) case of $\gamma=2$, 
where the series exactly truncates at cubic order, 
\begin{eqnarray}\label{Lig2}
& \delta\mathscr{L}_{I}=\frac{1}{2}\frac{\rho_0}{c_{s0}^2}\left(\delta\dot{\Phi}+{\bm v}_0\cdot\nabla\delta\Phi\right)(\nabla\delta\Phi)^2 \quad (\mbox{BEC}). 
\end{eqnarray}
\section{Effective space-time metric}
\subsection{General equation of motion for phase perturbations}

The equation of motion for $\delta\Phi$ can be found from \eqref{LD2} as follows
\begin{widetext}
\begin{multline}
\partial_\mu (f^{\mu\nu}_0\partial_\nu\delta\Phi)+\partial_t\left(-\frac{\rho_0}{c_{s0}^2}\frac 12(\nabla\delta\Phi)^2\right)
+\nabla\cdot\left(-\frac{\rho_0{\bm v}_0}{c_{s0}^2}\frac 12(\nabla\delta\Phi)^2\right)
+(2-\gamma)\frac{1}{c_{s0}^2}\left(\delta\dot{\Phi}+{\bm v}_0\cdot\nabla\delta\Phi\right)\nabla\cdot(\rho_0\nabla\delta\Phi)\\ 
+\nabla\cdot\left(-\frac{\rho_0}{c_{s0}^2}(\delta\dot{\Phi}+{\bm v}_0\cdot\nabla\delta\Phi)\nabla\delta\Phi\right)
+\frac{(2-\gamma)}{2c_{s0}^2}(\nabla\delta\Phi)^2\nabla\cdot(\rho_0\nabla\delta\Phi)+\nabla\cdot\left(-\frac{\rho_0}{2c_{s0}^2}(\nabla\delta\Phi)^2\nabla\delta\Phi\right)
=0.\label{eomphi}
\end{multline}
\end{widetext}
The above equation is the basic underlying nonlinear wave equation for $\delta\Phi$ 
on which the following considerations will be based. 
We note there that even though the Lagrangian of Eq.~\eqref{LD2} is an infinite series 
(for $\gamma \neq 2$) in the interaction term, the equation of motion for $\delta\Phi$ terminates at cubic order. 
Furthermore we observe that when we take nonlinearity into account, evidently the wave 
equation for $\delta\Phi$ 
cannot be brought into the form of a (massless) Klein-Gordon (KG) wave equation. Difference between any two solutions of the fluid equations, Eq.~\eqref{continuity}-Eq.~\eqref{euler} originating from two different boundary conditions can be expressible in terms $\delta\Phi$ satisfying Eq.~\eqref{eomphi}.
 
\subsection{On the choice of background}

Using the decomposition $\rho=\rho_{0}+\delta\rho$, ${\bm v}={\bm v}_{0}+\delta {\bm v}$ 
of the solution $\rho$, ${\bm v}$ of the fundamental fluid-dynamical equations, it is not always possible to physically distinguish a background ($\rho_0, {\bm v}_0$), and identify it uniquely.  For concreteness, to illustrate this,  
say we consider an ideal gas in a box, for which $\rho_0$ is constant everywhere.
One of the walls of the box can be moved by a piston to introduce perturbations.  We then increase the pressure in the box adiabatically by pushing the piston. 
After equilibrium has been reached, a pressure change obtains everywhere (as a result of a polytropic equation of state), and the change in velocity is zero, so that $\nabla\delta\Phi=0$, where
$\delta\Phi= C t$, $C$ a constant, can be found from Eq.~\eqref{10}.  
In Eq.~\eqref{eomphi}, all the nonlinear parts involve $\nabla\delta\Phi$, therefore $\delta\Phi=Ct$, which satisfies $\partial_\mu (f^{\mu\nu}_0\partial_\nu\delta\Phi)=0$, which is the usual wave equation in a static uniform background. Such a wave equation has a general solution of the form $f(x-c_{s0}t)+g(x+c_{s0}t)$,
where $f$ and $g$ are two well behaved functions, and the phase perturbation is $\delta\Phi=Ct=-A~(x-c_{s0}t)+A~(x+c_{s0}t)$ where $A$ is another constant, satisfying $Ac_{s0}=C/2$. 
However, this  perturbation $\delta\Phi$  is achievable from an {\it infinite number of backgrounds} which we would start from. 

We see from this simple example that, 
formally, it is certainly correct to write any solution of the fluid equations as perturbations on top of a background, but it is not always physically possible and meaningful to 
uniquely identify perturbations and background. Conversely, if we imagine the piston executes a small-amplitude  oscillatory motion, the separation into perturbation and background is meaningful. 
Thus an arbitrary solution of fluid equations has a {\it physical} background and perturbations on top of it only when the solution can be separated into two parts; one having slow variations in space and time, i.e., the background, and the other having relatively fast variations in space and time, i.e., the perturbation (the sound wave). 

\subsection{Linearized regime}
Only when we are operating in the linearized in $\delta\Phi$  regime of \eqref{eomphi}, the equation of motion for $\delta\Phi$ 
can be put into the form of the massless KG equation. Then one may readily define the effective metric from $f^{\mu\nu}_0$. In 3+1D, it reads
\begin{equation}\label{gmnl}
g_{\mu\nu}\coloneqq\frac{\rho_0}{c_{s0}}\begin{bmatrix}
 -(c_{s0}^{2}-v_0^{2}) & \vdots & -{\bf v}_0^T\\
\cdots&\cdots&\cdots\cdots \\
-{\bf v}_0&\vdots &\mathbb{I}
\end{bmatrix}
\end{equation}
which corresponds to the conventional acoustic metric of the standard paradigm
of analogue gravity.

\subsection{Redefining the space-time metric taking nonlinearity into account}

If the solution of Eq.~\eqref{eomphi} is known, the
density follows from Eq.~\eqref{10}, and the velocity is  $\bm{v}=\bm{v}_0+\nabla\delta\Phi$. 
The difference between two solutions of the fluid-dynamical equations originating from two different boundary value problems imposed on the system can always be expressed by a single scalar which is $\delta\Phi$. We can then choose to define a background solution of the fluid equations.  
We may define a new $\rm{\mathfrak{background}\,\,\mathfrak{(ii)}}$ (see table \ref{backgroundtable}) 
by $\rho_{(0)} =\rho_0+\delta\rho, \bm {v}_{(0)}=\bm{v}_0+\nabla\delta\Phi$, where
$\delta\Phi$ is a (sufficiently slowly varying) solution of the full nonlinear \eqref{eomphi}.   
 The fluid equations linearized  over this new background give a wave equation 
 for the first-order perturbation $\Phi_{(1)}$ which again takes the form of the massless KG equation. 
The new perturbation $\Phi_{(1)}$ is not aware of the original background (i), it couples to the new 
$\rm{\mathfrak{background}\,\,\mathfrak{(ii)}}$ in a similar manner to what is postulated in the scalar theory of gravity proposed in  \cite{Novello_2013}, and as in other field theories of gravity over 
curved or Minkowski background space-times \cite{feynman2018feynman,Sasha,Deser_1987,Gupta}. 
Therefore by linearizing with respect to the $\rm{\mathfrak{background}\,\,\mathfrak{(ii)}}$ ,  
we have 
\begin{equation}
\partial_\mu (f^{\mu\nu}_{(0)}\partial_\nu\Phi_{(1)})=0,
\end{equation}
\begin{equation}\label{fmn}
f^{\mu\nu}_{(0)}\coloneqq\frac{\rho_{(0)}}{c_{s(0)}^{2}}\begin{bmatrix}
-1 & \vdots & -{\bf v}_{(0)}^T \\
\cdots&\cdots&\cdots\cdots \\
-{\bf v}_{(0)}&\vdots & c_{s(0)}^{2}\mathbb{I}-{\bf v}_{(0)}\otimes{\bf v}_{(0)}
\end{bmatrix}.
\smallskip
\end{equation}
The acoustic metric then is again of the form of \eqref{gmnl}, replacing $0\rightarrow (0)$, 
\begin{equation}
{\mathfrak g}_{\mu\nu}\coloneqq\frac{\rho_{(0)}}{c_{s(0)}}\begin{bmatrix}
 -(c_{s(0)}^{2}-v_{(0)}^{2}) & \vdots & -{\bf v}_{(0)}^T \\
\cdots&\cdots&\cdots\cdots \\
-{\bf v}_{(0)}&\vdots &\mathbb{I}
\end{bmatrix}.
\end{equation}

The spacet-time metrics associated to the two backgrounds are generally related by 
\begin{equation}
{\mathfrak g}_{\mu\nu}=g_{\mu\nu}+h_{\mu\nu}, 
\end{equation}
where the $h_{\mu\nu}$ represent the difference in acoustic metrics 
between $\rm{\mathfrak{background}\,\,\mathfrak{(ii)}}$ and background (i), and can be expressed as functions 
of $\partial_\mu\delta\Phi$. Therefore, here gravity can be generated from a single self-interacting scalar over an arbitrary background, cf.~\cite{Sasha, Deser_1987}. Expanding  up to second 
order in $\delta\Phi$, we obtain 
\begin{widetext}
\begin{multline}
h_{tt}=\frac{\rho_0}{c_{s0}}\left[\left(\frac{\gamma +1}{2}-\frac{(3-\gamma)}{2}\frac{v_0^2}{c_{s0}^2}\right)\delta\dot{\Phi}
+{\bm v}_0\cdot \nabla\delta\Phi\left(\frac{5+\gamma}{2}-\frac{(3-\gamma)}{2}\frac{v_0^2}{c_{s0}^2}\right)\right]\\ 
 -\frac{\rho _0}{c_{s0}}\left[\delta\dot{\Phi}^2\frac{(3-\gamma)}{2c_{s0}^2}\left\{(\gamma -1)+\left(1-\frac{v_0^2}{c_{s0}^2}\right)\frac{(5-3\gamma)}{4}\right\}
+(\delta\dot{\Phi}{\bm v}_0\cdot\nabla\delta\Phi)\frac{(3-\gamma)}{2c_{s0}^2}\left\{2\gamma+\left(1-\frac{v_0^2}{c_{s0}^2}\right)\frac{(5-3\gamma)}{2}\right\}\right]
\\
  +\frac{\rho _0}{c_{s0}}\left[(\nabla\delta\Phi)^2\left\{\frac{(\gamma +1)}{2}+\left(1-\frac{v_0^2}{c_{s0}^2}\right)\frac{(3-\gamma)}{4}\right\}
-({\bm v}_0\cdot \nabla\delta\Phi)^2\frac{(3-\gamma)}{2c_{s0}^2}\left\{(\gamma +1)+\left(1-\frac{v_0^2}{c_{s0}^2}\right)\frac{(5-3\gamma)}{4}\right\}\right],
\end{multline}
\begin{multline}
h_{ti}=\frac{\rho_0}{c_{s0}}\left[\left\{v_0^i\frac{(3-\gamma)}{2c_{s0}^2}\left(\delta\dot{\Phi}+{\bm v}_0\cdot\nabla\delta\Phi\right)-\partial_i\delta\Phi\right\}
+\left\{\frac{(3-\gamma)}{2c_{s0}^2}\left(\delta\dot{\Phi}+{\bm v}_0\cdot\nabla\delta\Phi\right)\partial_i\delta\Phi\right\}\right]\\
+\frac{\rho_0}{c_{s0}}v_0^i\frac{(3-\gamma)}{2c_{s0}^2}\left\{\frac 12(\nabla\delta\Phi)^2-\frac{(5-3\gamma)}{4c_{s0}^2}\left(\delta\dot{\Phi}+{\bm v}_0\cdot\nabla\delta\Phi\right)^2\right\},
\end{multline}
\begin{equation} \hspace*{-7em}
h_{ij}=\frac{\rho_0}{c_{s0}}\delta_{i,j}\left[-\frac{(3-\gamma)}{2c_{s0}^2}\left(\delta\dot{\Phi}+{\bm v}_0\cdot\nabla\delta\Phi\right)
+\frac{(3-\gamma)}{2c_{s0}^2}\left\{\frac 12(\nabla\delta\Phi)^2-\frac{(5-3\gamma)}{4c_{s0}^2}\left(\delta\dot{\Phi}+{\bm v}_0\cdot\nabla\delta\Phi\right)^2\right\}\right], 
\end{equation}
\end{widetext}
where $i,j = x,y,z$. 

We shall see in Sec. \ref{RIS} that nonlinearity comes into play over time for the simplest possible 
nonlinear wave in fluid. 
With progressing time, starting from a linear approximation, 
 $\delta\Phi$ reveals the  nonlinearity of \eqref{eomphi},
 as a consequence changing the proper definition of background, and produces the new metric $\mathfrak{g}_{\mu\nu}$. 

\begin{table}[h]
\begin{center}
\begin{tabular}{|c|c|}
\hline
Background {\rm (i)} & $\rm{\mathfrak{Background}\,\,\mathfrak{(ii)}}$ \\\hline 
$\delta \Phi$: linearized Eq.~\eqref{eomphi}
&~~ $\delta \Phi$: full nonlinear Eq.~\eqref{eomphi}
\\\hline
$\rho_0, \bm{v}_{0}$ & $\rho_{(0)}, \bm{v}_{(0)}$ \\\hline  
 $g_{\mu\nu}$ &   ${\mathfrak g}_{\mu\nu}$   \\
 \hline
\end{tabular}
\caption{Definition of background variables and their associated metrics. 
In Background (i) , the velocity potential perturbation $\delta \Phi$ represents a massless scalar field, whereas in 
$\rm{\mathfrak{Background}\,\,\mathfrak{(ii)}}$  it satisfies the full nonlinear wave equation \eqref{eomphi}.}
\label{backgroundtable}
\end{center}
\end{table}

\subsection{Wave equation for time independent backgrounds}
If the background (i) is time independent, the equations satisfied by $\delta\Phi$ 
have time translation symmetry; $t\rightarrow t'=t-\epsilon$, $\delta\Phi\rightarrow\delta\Phi-\epsilon\delta\zeta$, $\partial_t\delta\Phi=-\delta\zeta$, where $\delta\zeta$ is Bernoulli's function. Substituting the new time-translated $\delta\Phi$ in Eq.~\eqref{dh} yields 
\begin{equation}\label{dh2}
-\frac{\rho}{c_s^2}\partial_t\delta\zeta-\frac{\rho}{c_s^2}{\bm v}\cdot\nabla\delta\zeta+\partial_t\delta\rho=0.
\end{equation}
Taking another partial time derivative of Eq.~\eqref{dh2}, and using the continuity equation, we find
\begin{equation} \label{dzeta}
\partial_\mu (\sqrt{-{\mathfrak g}} {\mathfrak g}^{\mu\nu}\partial_\nu\delta\zeta)=0,
\end{equation}
where $\mathfrak g$ is the determinant of ${\mathfrak g} _{\mu\nu}$.  
Equation \eqref{dh2}, and as a consequence, Eq.~\eqref{dzeta} are 
valid 
without the requirement of enthalpy $h$ being in the specific form (the case of polytropic equation of state)) discussed after Eq.~\eqref{h}. 
Nevertheless, Eq.~\eqref{dzeta} can also be found from the equation of motion of $\delta\Phi$
if we consider the background (i) as a uniform stationary medium. 
Thus here the time derivative $\delta\dot{\Phi}=-\delta\zeta$, instead of $\delta\Phi$ itself, behaves like a massless scalar field over a curved space-time, where the nonlinear self-interaction of $\delta\Phi$ is responsible for generating $h_{\mu\nu}$ in addition to the original Minkowski background.




\section{Riemann wave equation and Riemann invariants}\label{RIS}


We now treat a case where analytical techniques to study nonlinear sound are well established. 
We consider a one-dimensional sound wave propagating in a uniform static medium, i.e., a  
background of type  (i) is our starting point. Therefore, from Eq.~\eqref{eomphi}, we have 
\begin{eqnarray}\label{eomphi2}
&[-\partial_t^2+c_{s0}^2\partial_x^2]\delta\Phi=2(\partial_x\partial_t\delta\Phi )\partial_x\delta\Phi +(\gamma-1)(\partial_x^2\delta\Phi)\partial_t\delta\Phi\nonumber\\ 
& +\frac{(\gamma+1)}{2}(\partial_x \delta\Phi)^2 \partial_x^2\delta\Phi .
\end{eqnarray}


Instead of directly starting from this equation for $\delta\Phi$, we reinstate the problem in terms of
Riemann invariants, the powerful method being due to the seminal paper of Riemann in 
1860 \cite{Riemann1860},  
solving the problem of 1D shock waves analytically. 
The fluid equations in one spatial dimension lead to the 
Riemann invariants being given by the partial differential equations \cite{Landau1987Fluid}:
\begin{eqnarray}
& \left[\frac{\partial}{\partial t}+(v+c_s)\frac{\partial}{\partial x}\right]J_+=0,\label{RI}\\
& \left[\frac{\partial}{\partial t}+(v-c_s)\frac{\partial}{\partial x}\right]J_-=0,\label{RI2} 
\end{eqnarray}
The  total sound speed is $c_s=c_{s0}+\delta c_s$, and the invariants 
can be expanded $J_{\pm}=J_{0\pm}+\delta J_{\pm}=v\pm \int\frac{dp}{\rho c_s}$. 
In a polytropic medium, we have \cite{Landau1987Fluid}
\begin{eqnarray}
& J_{0\pm}=\pm\frac{2c_{s0}}{\gamma -1},\nn
& v=\frac{\delta J_++\delta J_-}{2}=\frac{J_++J_-}{2},\nn
& \rho=\rho_0\left\{\frac{(\gamma -1)}{4c_{s0}}(J_+-J_-)\right\}^{\frac{2}{\gamma -1}}.
\label{JEqs}
\end{eqnarray}
\begin{figure}[t]
  \centering
 
 \includegraphics[scale=0.3]{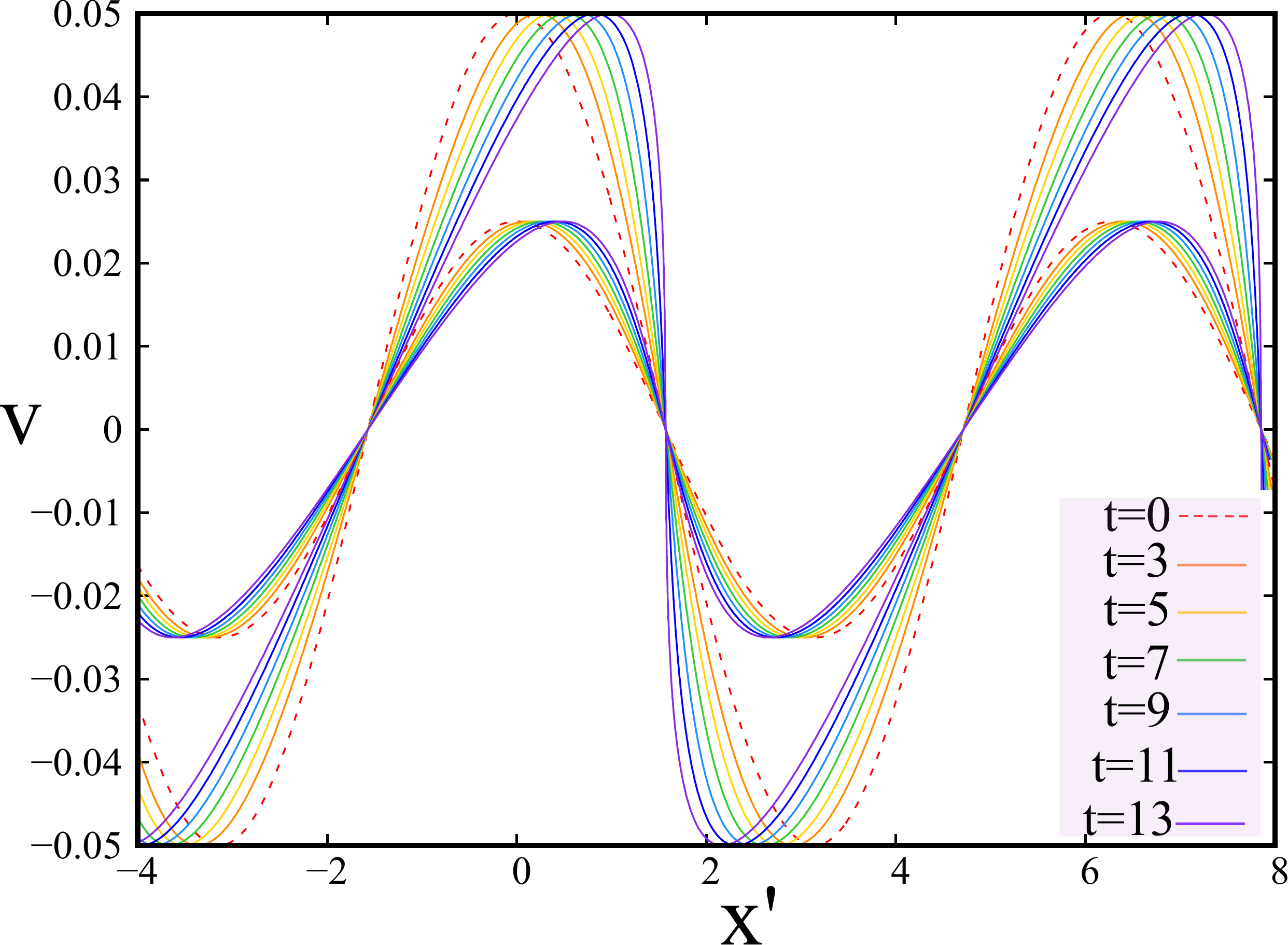}
   \caption{Simple wave solution of the Riemann wave equation \eqref{RW}, displayed  
   in the comoving frame $x'$ frame, $x'=x-c_{s0}t$, for wave profile with initial configuration: $v=A\cos kx$ at $t=0$; $A=0.05$ and $A=0.025$; and $k=1,~\hbar=1$ in units of $c_{s0}$, $c_{s0}$ is set to unity. 
    For a BEC, $\gamma=2$. 
    According to the linearized solution from Eq.~\eqref{vxl}, the wave does not change shape over time, the red dotted lines are the solutions for all $t$. Due to nonlinearity, points of larger velocity $v$ move with higher speed, resulting in a change in shape of the profile as displayed, where we consider times  $t<t_{\rm shock}$, where $t_{\rm shock}=40/3$ for the cosine wave with amplitude $A=0.05$, and $t_{\rm shock}=80/3$ for the cosine wave with $A=0.025$. The degree of nonlinearity is different for the two waves at the same instant; e.g., at $t=13$, the profile with $A=0.05$ is closer to approach the shock than that with $A=0.025$.   
  }
\label{fig1}
\vspace*{-0.5em}
\end{figure}
\subsection{Simple wave solution} 
Now, we consider the simplest possible case, i.e., a wave traveling in a particular direction, called a 
{\it simple wave} \cite{Landau1987Fluid}. Then, $J_-$ is constant everywhere ($\delta J_-=0 
\rightarrow J_-=J_{0-}$), and the variation of $J_+$ represents the propagation of the Riemann wave. 

Constancy of $J_-$ throughout the whole $x,t$ domain implies, from Eqs.~\eqref{JEqs} 
\begin{eqnarray}
\label{rhov}
\rho=\rho_{0}\left[1+\left(\frac{\gamma -1}{2}\right)\frac{v}{c_{s0}}\right]^{\frac{2}{\gamma-1}}.
\end{eqnarray}
Hence, a simple wave can be described by a single variable $v$ or $\rho$, 
The equation \eqref{RI} for $J_+$ gives the Riemann wave equation \cite{Landau1987Fluid}:
\begin{equation}\label{RW}
\frac{\partial v}{\partial t}+\left[c_{s0}+\left(\frac{\gamma +1}{2}\right) v\right]\frac{\partial v}{\partial x}=0. 
\end{equation}
This equation  has an analytic solution  \cite{Landau1987Fluid} 
\begin{eqnarray}
& v=f(\xi), \label{vx}\\
& x=\left[c_{s0}+\left(\frac{\gamma+1}{2}\right) v\right]t+\xi.\label{chh}
\end{eqnarray}
The above relation between $\xi$ and $x,~t$ represents a  
simple wave solution traveling along positive $x$ axis. 
In the linearized limit, the Riemann wave equation reduces to
\begin{equation}\label{RWl}
\frac{\partial v}{\partial t}+c_{s0} \frac{\partial v}{\partial x}=0. 
\end{equation}
Using again the method of characteristics gives the solution in the form
\begin{eqnarray}
& v=f(\xi), \label{vxl}\\
& x=c_{s0}t+\xi.
\end{eqnarray}
The linearized solution and the nonlinear solution are depicted in the Fig. \ref{fig1},
which depicts how the solution deviates from the linearized solution over time. 
Initially, the linearized description affords a sufficiently accurate description and 
the wave corresponds in the analogue gravity context to 
a massless scalar field over flat Minkowski space-time. However, with advancing 
time, that is as $t$ approaches $t_{\rm shock}$, this does not hold anymore, requiring the background to be redefined as a $\rm{\mathfrak{background}}$ of type
$\rm{\mathfrak{(ii)}}$, which is represented by the 
Riemann wave itself, with the metric
${\mathfrak g_{\mu\nu}}$. 
Thus the field $\delta\Phi$, satisfying Eq.~\eqref{eomphi2}, changes the background and 
effectively creates gravity, that is a curved space-time, from a Minkowski space-time. 

The shock time $t_{\rm shock}$ is defined as the instant 
when at the shock location $x=x_{\rm shock}$, $\frac{\partial v}{\partial x}$ goes to infinity. 
This can be analytically shown from the method of characteristics to solve the Riemann wave equation \cite{Landau1987Fluid}. For a simple wave with initial profile (at t=0) $v=A \cos kx$, $t_{\rm shock}=\frac{2}{(\gamma +1)Ak}$ \cite{Papon}.

Inspecting Fig.~\ref{fig1}, one may choose an instant $t_{\rm lin}$  
which is setting an upper limit in time until which the solution of the Riemann wave equation 
can be approximately regarded as residing in a linearized regime. 
The choice of $t_{\rm lin}$ depends on the required precision of reproducing the exact solution
while still staying in that linearized regime.
If the observation time in any experiment fulfills $t_{\rm obs}<t_{\rm lin}$, then the solution of the Riemann wave equation can be considered as a massless field over the background (i), with 
metric (absorbing a constant conformal factor $\rho_0/c_{s0}$). 
Writing the originally 3+1D 
metric  in a quasi-1D system in 1+1D form , we have \footnote{See for the derivation of the 1+1D metric from the embedding 
 3+1D metric Ref.~\cite{Papon}},   
\begin{equation}
g_{\mu\nu}\coloneqq
\begin{bmatrix}
 -c_{s0}^{2} & 0 
 \\
0 & 1 
\end{bmatrix}.
\end{equation}
On the other hand, if $t_{\rm obs}\gtrsim t_{\rm lin}$,  
there are two possible procedures. 
As a first option (1), for $t<t_{\rm lin}$, the Riemann wave is considered as a massless scalar field; for $t\geq t_{\rm lin}$, 
background (i) is reverted to ${\rm \mathfrak{background}}$ (ii), and one defines a new perturbation 
$\Phi_{(1)}$, which again varies faster (in space and time) than the Riemann wave.
The background is redefined, the Riemann wave is itself the background, leading to the 
new acoustic 1+1D metric of type $\rm{\mathfrak{(ii)}}$, again taking over the conformal
factors from the 3+1D embedding space of a quasi-1D system \cite{Papon},  
\begin{equation}
\mathfrak{g}_{\mu\nu}({\bm{r}},t)\coloneqq 
\frac{c_{s0}}{\rho_0} 
\frac{\rho_{(0)}}{c_{s(0)}}
\begin{bmatrix}
 -(c_{s(0)}^{2}-v_{(0)}^{2}) & -v_{(0)} 
 \\
-v_{(0)} & 1 
\end{bmatrix}.
\end{equation}
As a second option (2) one can instead consider the Riemann wave as the background from the very beginning ($t=0$), and linearize the fluid equations to find the corresponding massless scalar field, $\Phi_{(1)}$.
That linear perturbation $\Phi_{(1)}$ again can behave nonlinearly in some domain ${\bm x}-t$, therefore one needs to redefine the background again.

\subsection{Metric components due to nonlinearity}
The new acoustic 1+1D  metric can be decomposed as follows   
 \begin{eqnarray}\label{ngw}
 \mathfrak{g}_{\mu\nu}(x,t)&=&g_{\mu\nu}+h_{\mu\nu} \nn
& = & \begin{bmatrix}
-c_{s0}^2 & 0 \\
0 & 1
\end{bmatrix}
 +h_{\mu\nu}(x,t),\quad (\mu,\nu = t,x). 
 \end{eqnarray}
The metric perturbations $h_{\mu\nu}(x,t)$ 
are  to second order given by \cite{Papon}
\begin{eqnarray}
h_{tt}&=&-c_{s0}^2\!\!\left[\!\frac{(\gamma +1)}{2}\frac{\delta\rho}{\rho_{0}}\!+\!\frac{(\gamma^2-1)}{8}\!\left(\frac{\delta\rho}{\rho_{0}}\right)^2\right]\!+v^2,\label{htt}\\
h_{tx}&=&h_{xt}=-v\left(1+\frac{(3-\gamma)}{2}\frac{\delta\rho}{\rho_{0}}\right),\\
h_{xx}&=&\frac{(3-\gamma)}{2}\frac{\delta\rho}{\rho_{0}}-\frac{(3-\gamma)(\gamma -1)}{8}\left(\frac{\delta\rho}{\rho_{0}}\right)^2. \label{hxx}
 \end{eqnarray}
The $h_{\mu\nu}$ are expressible in terms of a single variable $v$ from Eq.~\eqref{rhov}.

The perturbations on top of the background {\it back-react} 
therefore on the definition of the background and the resulting acoustic metric. 
We limit ourselves to the nondispersive (Thomas-Fermi) limit of negligible density
variations. As the wave slopes in Fig. \ref{fig1} become steeper and steeper with time, the Thomas-Fermi assumption breaks down. 
We may take this limitation into account by imposing $t\lesssim t_{\rm shock}$.

\section{Source tensor of the effective gravitational field} 
\label{source terms section}

From Eqs.~\eqref{ngw}-\eqref{hxx}, and using Eq.~\eqref{rhov}, we find the metric components $h_{\mu\nu}=h_{\mu\nu}(v)$.
We observe the Riemann Eq.~\eqref{RW} remains true for any analytical function of $v=v(x,t)$: 
Given any such function $\mathscr{F}(v)$, for simple waves along positive $x$ axis, we thus have $\frac{\partial\mathscr{F}(v)}{\partial t}+\left(c_{s0}+\frac{\gamma+1}2 v\right)\frac{\partial \mathscr{F}(v)}{\partial x}=0$. 
Hence, from Eq.~\eqref{RW}, we have the following relation:
$\frac{\partial h_{\mu\nu}}{\partial t}+\left(c_{s0}+\frac{\gamma+1}2 v\right)\frac{\partial h_{\mu\nu}}{\partial x}=0$. 
We introduce $\Box_+=\left(\frac{1}{c_{s0}}
\frac{\partial}{\partial t}+\frac{\partial}{\partial x}\right)$, $\Box_-=-
\frac{1}{c_{s0}}
\frac{\partial}{\partial t}+\frac{\partial}{\partial x}$ and 
thus have $\Box=\Box_-\Box_+=\Box_+\Box_-=-\frac{1}{c_{s0}^2}\frac{\partial^2}{\partial t^2}+\frac{\partial^2}{\partial x^2}$. 
Therefore, we obtain the wave equations 
\begin{equation}
 \Box h_{\mu\nu}= 
 S_{\mu\nu}, \label{htxhtt}
\end{equation} 
where the {\it nonlinear source term} is given by, 
\begin{multline}
\hspace*{-1em}S_{\mu\nu}=
\frac{\gamma+1}{2c_{s0}^2}(\partial_xh_{\mu\nu})(\partial_t-c_{s0}\partial_x)\!((2-\gamma)h_{tx}+2c_{s0}h_{xx})\\
+\frac{\gamma+1}{2c_{s0}^2}((2-\gamma)h_{tx}+2c_{s0}h_{xx})(-c_{s0}\partial_x^2h_{\mu\nu}+\partial_x\partial_t h_{\mu\nu}). 
\end{multline} 
Note that this source term 
 appears if and only if the nonlinearity in the Riemann wave equation \eqref{RW} is present, 
 and hence vanishes in the limit of linearized acoustics.
 In Appendix \ref{non-simple}, we derive the source term of the wave equation \eqref{htxhtt}
for the more general case of a non-simple wave.

In general relativity,  the right-hand side of \eqref{htxhtt} contains 
the gravitational Landau-Lifshitz (LL) energy-momentum pseudo-tensor 
$t_{\mu\nu}$, in the form  
$t_{\mu\nu}-\frac{1}{2}\eta_{\mu\nu}t^\lambda_{~~\lambda}$ in traceless-transverse gauge, with $\eta_{\mu\nu}={\rm diag} (-1,1,1,1)$  \cite{weinberg1972gravitation}. 
The proper gravitational source term and its analogue model counterpart share some common properties, but there are also notable differences. 
Both in gravity proper and within our analogue model they are quadratic in $h_{\mu\nu}$ and contain its first and second order 
space-time coordinate derivatives, and the components of the GW act as a source themselves  \cite{weinberg1972gravitation}.
The presence of the lab frame, with absolute Newtonian time $t$, however engenders that the (coordinate reparametrization) general 
covariance property of general relativity is not reflected in (so far existing) analogue models 
\footnote{See, e.g., the discussion of diffeomorphism invariance contained in the review \cite{BLV}}.  
The energy-momentum conservation law from the Bianchi identities 
thus does not hold in the analogue model. 

One may contrast this with the tensor describing conserved  energy and momentum 
canonically  derived from the Lagrangian
density $ \delta\mathscr{L}$ as 
$T^\mu_{~~\nu}=\frac{\partial \delta\mathscr{L}}{\partial (\partial_\mu \delta\Phi)}\partial_\nu\delta\Phi-\delta^\mu_\nu \delta\mathscr{L}$, which can be defined for a uniform and stationary background. 
Considering up to cubic terms of $\delta\Phi$ in $\delta\mathscr{L}_I$, 
one has for the interaction part of the Lagrangian density, from \eqref{LD},  
\begin{eqnarray}
\label{L3} 
\delta\mathscr{L}_{I}&=&\frac{\rho_0}{2c_{s0}^2}\left(\delta\dot{\Phi}+{\bm v}_0\cdot\nabla\delta\Phi\right)
\nn
& & \times 
\left[(\nabla\delta\Phi)^2-\frac{(2-\gamma)}{3c_{s0}^2} \left(\delta\dot{\Phi}+{\bm v}_0\cdot\nabla\delta\Phi\right)^2 \right] \nn & &+ \ord([\delta\Phi]^4).  
\end{eqnarray}
Then one obtains for the canonical energy-momentum tensor, to $\ord([\delta\Phi]^3)$ 
\begin{equation}
T^t_{~~t}=\frac{\rho_0}{2c_{s0}^2}\left(\delta\dot\Phi^2+c_{s0}^2(\nabla\delta\Phi)^2\right)-\frac{(2-\gamma)\rho_0}{3c_{s0}^4}\delta\dot\Phi^3,
\end{equation}
\begin{equation}
T^t_{~~i}=\frac{\rho_0}{c_{s0}^2}\delta\dot\Phi\partial_i\delta\Phi+\frac{\rho _0}{2c_{s0}^2}\left[(\nabla\delta\Phi)^2-\frac{(2-\gamma)}{c_{s0}^2}\delta\dot\Phi^2\right]\partial_i\delta\Phi,
\end{equation}
\begin{equation}
T^i_{~~t}=\left[\frac{\rho_0}{c_{s0}^2}\delta\dot\Phi\partial_i\delta\Phi-\rho_0\partial_i\delta\Phi\right]\delta\dot\Phi,
\end{equation}
\begin{equation}
T^i_{~~j}=\left[\frac{\rho_0}{c_{s0}^2}\delta\dot\Phi\partial_i\delta\Phi-\rho_0\partial_i\delta\Phi\right]\partial_j\delta\Phi-\delta^i_{~~j}\delta\mathscr{L}.
\end{equation}
One readily verifies that there is no one-to-one correspondence between $T_{\mu\nu}$ and 
$S_{\mu\nu}$. The latter involves derivatives of density and velocity perturbations (gradients
of the $h_{\mu\nu}$), while $T_{\mu\nu}$ contains only algebraic functions of the $h_{\mu\nu}$. 

\section{Experimental considerations} 

\subsection{Shock times $t_{\rm shock}$}

Here we show that the  nonlinearity after imprinting a wave profile becomes manifest 
for typical BEC parameters, e.g., in $^{87}\!$Rb  on time scales much less than their lifetime, also and in particular 
for realized analogue gravity setups \cite{Munoz}.  

For a BEC, the time after which the shock singularity is reached after initially imprinting 
a cosine profile for the velocity is given by  
\begin{equation}
t_{\rm shock}=\frac2{3Ak} \qquad \mbox{(BEC)}. 
\end{equation}
We specify, setting $\hbar =m =1$, the wave vector  $k$ in units of $1/\xi_c \coloneqq c_s$ 
and $A$ also in units of $c_s$ or $1/\xi_c$. 
Furthermore,  $\mu= 1/\xi_c^2=c_s^2$,  which means that $t_{\rm shock}$ can be expressed in units of $1/\mu$. 

We should have as a minimal requirement for nonlinearity to be observable that 
$t_{\rm shock} \ll \tau$, the lifetime  of the BEC (which is mainly limited 
 by three-body recombination).  
 We assume that the laser wavelength for phase imprinting (also see below subsection)
is scaled in units of $2\pi \xi_c$, and $k$ expressed in units of $1/\xi_c=c_{s0})$, so that $t_{\rm shock}[1/\mu]\sim \frac2{3A[c_{s0}]k[1/\xi_c]}$. 
 In Ref.~\cite{Soeding}, the chemical potential is $\mu\sim 2$\,kHz [100\,nK]. 
 With $A=0.05$ used in Fig. \ref{fig1}, for $k=1$, then, returning to 
 dimensionful units via $t_{\rm shock}[1/\mu] = t_{\rm shock}[1/2\pi \mu$\,[Hz]], we have 
  $t_{\rm shock}\simeq 1.1 $\,msec ($\simeq 2.2$\,msec for $A=0.025$),  which is much less than the lifetime of order seconds which was observed in Ref.~\cite{Soeding}.
 In the $^{87}\!$Rb 
 analogue black hole experiment \cite{Munoz}, which provided an observation of analogue Hawking radiation, the chemical potential as defined from 
 $\xi_c$ is much less, of order $\mu\sim$\,30 Hz [1.4 nK]. 
To derive this $\mu$ from $1/\xi_c^2$,  
we use as the coherence length the geometrically averaged quantity defined in \cite{Munoz} 
  (averaged between upstream and downstream regions relative to the horizon) 
which is $\xi_c\sim 1.8\,\mu$m. The Hawking temperature in \cite{Munoz} 
 is $T_{\rm H} \sim \mu/4$. 
Setting $k=1/4$ $(\lambda =8\pi)$,  which is of order the wave vector of the dominant 
Hawking modes (in the infrared), 
we then have 
$t_{\rm shock}\sim 0.57$\,sec for $A=0.025$, choosing here a value for the velocity perturbation amplitude which is order-of-magnitude consistent with the density-density Hawking correlations measured in \cite{Munoz}. 
This value for $t_{\rm shock}$ is 
 still much less than the lifetime of the experiment, which strongly increases due to the much lower densities used in the $^{87}\!$Rb  experiment  of \cite{Munoz} when compared to that of 
 \cite{Soeding}.
 We also note that shock waves in a BEC have in fact been observed, e.g., in \cite{meppelink2009observation}, with the theory developed in \cite{Damski2004}.  
 Finally, similar considerations can be performed for fluids of light, in which shock wave dynamics 
 has been observed and analyzed as well \cite{Bienaime}.
 
While the above estimates for $t_{\rm shock}$ were derived for 
a strictly one-dimensional flow which obeys the Riemann wave equation, using parameters
from previously conducted experiments on $^{87}\!$Rb which have not been conducted 
in quasi-one-dimensional setups (Ref.~\cite{Munoz} operates in the transition region to quasi-1D), 
we conclude that to consider the nonlinearity of the BEC fluid is in general unavoidable, 
also and in particular in typical analogue gravity setups.  
 
\subsection{Mass flux as a signature of nonlinearity}

\begin{figure}[tb]
  \centering
\vspace*{-2.5em}
\includegraphics[scale=0.195]{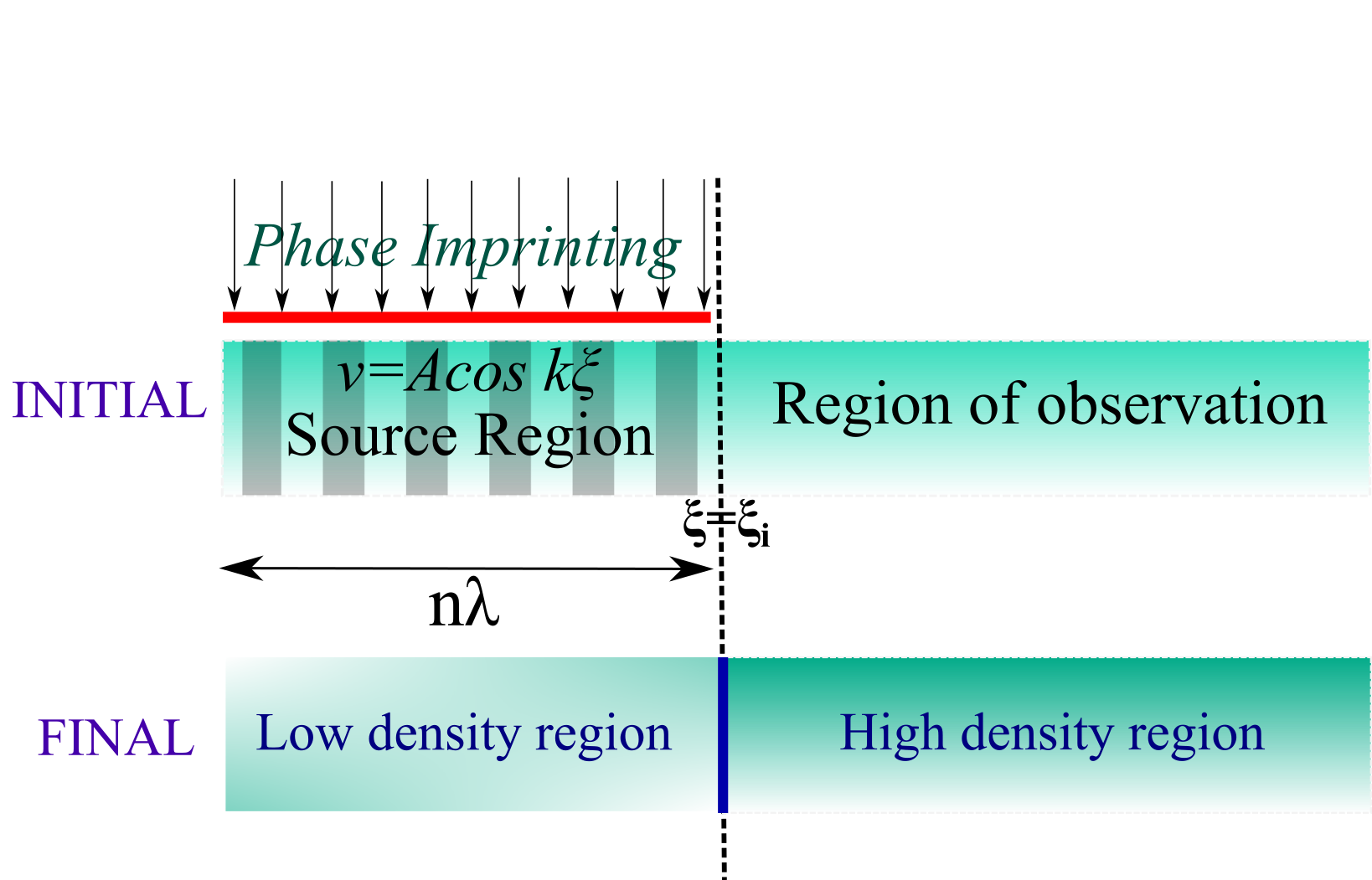}~
    \caption{
  Phase in the expression \eqref{PhiProfile} produces a monochromatic $v=A cos kx$. 
    Mass then flows  from source ($S$) to observation region ($RO$). 
    For the final equilibration stage, when the full wave train has entered the $RO$,  
    we separate $RO$ and $S$ 
    by a repulsive laser sheet barrier, 
    at time $n\lambda/c_{s0}$,  
    and absorption imaging determines the total mass displacement $\Delta M$ as a signature of nonlinearity.
    \label{figNonlinear}}
\end{figure}

To derive a simple experimental measure of nonlinearity, 
we compute in this subsection the time-averaged 
mass flux through $x=\xi_i$  (we use here background (i), see table \ref{backgroundtable}, left column).
We consider a homogeneous cloud of ultracold 
 $^{87}\!$Rb 
 in a cylindrical box trap, generated e.g. by the methods used in \cite{Gaunt},  
with radius $R\,(\ll \xi_c)$ and length $L\,(\gg \xi_c)$, 
choosing a region of length $n\lambda$
at the left end of the cylinder, denoted $S$ for source region  henceforth;
the observation region on the right 
is abbreviated $RO$, see~Fig.~\ref{figNonlinear}; here, $n\in  \mathbb{N}$.
We suggest to employ the phase imprinting technique, readily  
available in the quantum optical setup of ultracold gases
\cite{Denschlag,Leanhardt}. 
We create a spatial variation in the phase of the initial 
condensate wave function, within $S$, by red-detuned 
laser light turned on for a short duration $T$ (as short as to stay within the Raman-Nath regime of simple diffraction). 
The superfluid phase pattern $\Phi(x) \propto I(x)/\delta$, where
$I(x)$ is laser intensity and $\delta$ detuning from resonance, is then imprinted in $S$,  where 
\begin{equation}\label{PhiProfile} 
\Phi(x)=(A/k)\sin(kx)+C.
\end{equation}
Here, we put $A,C\geq 0$, and the constant $C$ is chosen such that $\Phi (x)>0$ within 
$S$ (red-detuning $\forall\, x\in S$). 
The thus created  bipartite 1D configuration, cf.~Fig.~\ref{figNonlinear},  produces simple waves \cite{Landau1987Fluid}.

We calculate the current averaged over time to be, see for details Appendix \ref{massflux}, 
\begin{equation}
\left< j \right >=\rho_{0}\frac{3-\gamma}{c_{s0}}\frac{A^2}{8}.\label{A^2} 
\end{equation}
For a BEC ($\gamma=2$), $\left< j \right > >0$,
indicating the mass amount 
$\Delta M = \pi R^2 \left< j \right > \Delta t $ with $\Delta t=2\pi n /(c_{s0}k)$ is 
flowing to RO from $S$.
The total mass displacement in \eqref{A^2} is proportional to $A^2\Delta t$ 
which is the hallmark of nonlinearity we aimed to derive.

\section{Conclusion and discussion}


At the very core of the analogue gravity concept is the separation
of the underlying field(s) into a background and small perturbations propagating on top of that background.
where the acoustic metirc components are functions of the background flow solution (density and velocity of the fluid medium). Here, we have tested this assumption of linearizing perturbations over a background flow. We find that generally, and worked out in detail for the simplest possible case of a one-dimensional wave over a uniform static background (corresponding to a Minkowski space-time for sound), that the assumption of linearity breaks down over the course of time. Beginning with a nonlinear perturbation over a background flow, we have demonstrated how the presence of nonlinearity in the perturbation equations of motion back-reacts on and thus changes the background flow, so that the acoustic metric is modified.


The phenomenon of emergence of a new metric due to nonlinearity of the underlying field(s), 
which arises 
naturally in analogue gravity, can be mapped to field theoretical formulations of gravity.
From a historical perspective, the {\it background field method} was introduced by Feyman, Deser,  and Gupta \cite{feynman2018feynman,Sasha, Deser_1987,Gupta} to quantize gravity. 
A space-time (classical background) is supposed to exist, and then the 
equations of motion for the metric perturbations are studied. 
In our case of analogue gravity in fluids, the metric perturbations 
(which change the  initial concept of background) 
are shown to be functions of nonlinear perturbations in the velocity scalar; hence 
gravity is a scalar field in the analogue context.
The idea of a scalar theory of gravity can be traced back to Newtonian gravity, where the gravitational potential satisfies Poisson's equation.  A generalization of the Newtonian gravitational potential within 
special relativity, initially proposed by Einstein and Grossmann \cite{einstein1987collected}, lacked general covariance (diffeomorphism invariance). More recently, a modern geometric scalar theory of gravity, respecting diffeomorphism invariance, has been proposed \cite{Novello_2013}, in which a nonlinear self-interacting field produces metric perturbations over Minkowski space-time. Matter fields do not perceive the latter space-time, they couple to the modified metric by the scalar field. 
We have demonstrated  
that in essentially the same way the self-interacting nonlinear terms in the perturbation of the scalar velocity potential are responsible for generating metric perturbations, and thus the acoustic metric is changed.
Introducing linear perturbations (on top of the new  $\rm{\mathfrak{background}}$ type
$\rm{\mathfrak{(ii)}}$), they do not interact with the Minkowskian background, and instead minimally couple to the modified acoustic metric ${\mathfrak g}_{\mu\nu}$, due to the nonlinear perturbation generated by the velocity potential scalar.
The idea of an emergence of a curved space-time metric from a Minkowski metric due to the nonlinearity of an underlying scalar field theory \cite{Novello_2011} can thus be tested with the established tools of fluid dynamics  in the context of analogue gravity.

Finally, we assumed in this work the nondispersive limit of fluid dynamics (which in BECs amounts to the Thomas-Fermi limit).  
Within the context of analogue gravity, the dispersive limit (that is, e.g., including the quantum pressure term in a BEC \cite{barcelo2001analogue}), has been studied under the banner of rainbow gravity 
\cite{CPPRainbow}.
Going beyond the nondispersive limit and hence approaching very closely the instant of the shock (which represents an analogue spacetime singularity) will be the subject of a future study.  


\section{Acknowledgments} 
This work has been supported by the National Research Foundation of Korea under 
Grants No.~2017R1A2A2A05001422 and 
No.~2020R1A2C2008103.

\begin{appendix}
\section{Non-simple waves in one spatial dimension}
\label{non-simple}
\subsection{Inhomogeneous wave equations for Riemann invariants}
We find from Eq.~\eqref{RI},
\begin{equation}\label{+-}
\Box_{\pm}\delta J_{\pm}=\left(\mp\frac{(\delta J_++\delta J_-)}{2c_{s0}}-\frac{(\gamma -1)}{4c_{s0}}(\delta J_+-\delta J_-)\right)\frac{\partial\delta J_\pm}{\partial x}.
\end{equation}
where $\Box_{\pm}=\pm\frac{1}{c_{s0}}\frac{\partial}{\partial t}+\frac{\partial}{\partial x}$.
Therefore, the  perturbation terms of Riemann invariants, $\delta J_{\pm}$ satisfy 
the following inhomogeneous wave equation
\begin{equation}
\Box \delta J_{\pm}=S_{\pm},
\end{equation}
where the source terms are 
\begin{eqnarray}\label{Spm}
& S_{\pm}=\frac{(\gamma +1)}{4c_{s0}^2}\left\{\delta J_\pm\frac{\partial^2\delta J_\pm}{\partial x\partial t}\mp c_{s0}\delta J_\pm\frac{\partial^2\delta J_\pm}{\partial x^2}\right\}\nonumber\\
&+\frac{(\gamma +1)}{4c_{s0}^2}\left\{\left(\frac{\partial \delta J_\pm}{\partial t}\right)\left(\frac{\partial \delta J_\pm}{\partial x}\right)\mp c_{s0}\left(\frac{\partial \delta J_\pm}{\partial x}\right)^2\right\}\nonumber\\
&-\frac{(\gamma -3)}{4c_{s0}^2}\left\{\delta J_\mp\frac{\partial^2\delta J_\pm}{\partial x\partial t}\mp c_{s0}\delta J_\mp\frac{\partial^2\delta J_\pm}{\partial x^2}\right\}\nonumber\\
&-\frac{(\gamma -3)}{4c_{s0}^2}\left\{\left(\frac{\partial \delta J_\mp}{\partial t}\right)\left(\frac{\partial \delta J_\pm}{\partial x}\right)\mp c_{s0}\left(\frac{\partial \delta J_\mp}{\partial x}\right)\left(\frac{\partial \delta J_\pm}{\partial x}\right)\right\}.\nonumber\\
\end{eqnarray}
with $\Box=\Box_-\Box_+=\left(-\frac{1}{c_{s0}^2}\frac{\partial}{\partial t^2}+\frac{\partial}{\partial x^2}\right)$.
\subsection{Source terms}
The expression \eqref{Spm} represents  an exact expression with source terms quadratic in $\delta J_\pm$.
The acoustic metric of $\rm{\mathfrak{background}}$ type  
$\rm{\mathfrak{(ii)}}$ is 
$\mathfrak{g}_{\mu\nu} (x,t)=\mathfrak{g}_{\mu\nu}(\rho, v)=\mathfrak{g}_{\mu\nu}(J_+,J_-)=\mathfrak{g}_{\mu\nu}(J_{0+}+\delta J_+,J_{0-}+\delta J_-)=\eta_{\mu\nu}+h_{\mu\nu}$, where $\eta_{\mu\nu}=\mathfrak{g}_{\mu\nu}(J_{0+},J_{0-})$, is the Minkowski metric with light speed replaced by the sound speed $c_{s0}$, and $h_{\mu\nu}$ consists of all the other terms in the Taylor series of $\mathfrak{g}_{\mu\nu}$. We then have $h_{\mu\nu}=h_{\mu\nu}(\delta J_+, \delta J_-)$.
Let us consider a general function $f=f(\delta J_+, \delta J_-)$. 
\begin{eqnarray}\label{bf}
& \Box f=g_+\Box \delta J_++g_-\Box\delta J_-+g_{++}(\Box_-\delta J_+)(\Box_+\delta J_+)\nonumber\\
&+g_{+-}(\Box_-\delta J_-)(\Box_+\delta J_+)+g_{-+}(\Box_-\delta J_+))(\Box_+\delta J_-)\nonumber\\
&+g_{--}(\Box_-\delta J_-)(\Box_+\delta J_-),
\end{eqnarray}
where $g_+=\frac{\partial f}{\partial \delta J_+},~g_-=\frac{\partial f}{\partial \delta J_-},~g_{++}=\frac{\partial^2 f}{\partial \delta J_+^2},~g_{+-}=\frac{\partial^2 f}{\partial \delta J_+\partial \delta J_-}=g_{-+},~g_{--}=\frac{\partial^2 f}{\partial \delta J_-^2}$. 
The first two terms in the right-hand side of Eq.~\eqref{bf} are at least quadratic in $\delta J_{\pm}$, whereas the second, third and fifth terms are at least cubic in  $\delta J_{\pm}$ according to Eq.~\eqref{+-}; and the fourth term is at least quadratic in  $\delta J_{\pm}$. 
We have
\begin{eqnarray}
& h_{xx}=\left(1+\frac{(\delta J_+-\delta J_-)}{\Delta J_0}\right)^\frac{3-\gamma}{\gamma -1}-1,\\
& h_{xt}=-\frac{(\delta J_++\delta J_-)}{2}(h_{xx}+1),\\
& h_{tt}=(1+h_{xx})(-c_{s0}^2(1+h_{xx})^{\frac{2(\gamma -1)}{(3-\gamma)}}+\frac{h_{xt}^2}{(1+h_{xx})^2})+c_{s0}^2,\label{nhtt}\nonumber\\
\end{eqnarray}
where $\Delta J_0=\frac{4c_{s0}}{\gamma -1}$. The number of independent components is two because $\delta J_+$ and $\delta J_-$ are independent, and $\delta J_{\pm}$ are derivable from 
the two Riemann-invariant equations. The perturbation terms of $J_{\pm}$ can be written in terms of $h_{xt}$ and $h_{xx}$,
\begin{equation}\label{jh}
\delta J_{\pm}=\pm\frac{\Delta J_0}{2}\left((1+h_{xx})^{\frac{\gamma -1}{3-\gamma}}-1\right)-\left(\frac{h_{xt}}{1+h_{xx}}\right).
\end{equation}
Therefore, using Eq.~\eqref{bf}, we can compute the source terms as a functions of the $h_{\mu\nu}$. Source terms $S_{\mu\nu}$ are functions of $h_{xx},~h_{xt}$ and all first order and second order partial derivatives of $h_{xt}$ and $h_{xx}$.
As an example, we first consider the case of simple waves \cite{Landau1987Fluid}. For a simple wave propagating along the positive $x$ axis, $\delta J_-$ is zero. The dynamics has only one degree of freedom corresponding to $\delta J_+$. This yields from the expression \eqref{jh},
\begin{equation}\label{hxthxx}
h_{xt}=-\frac{\Delta J_0}{2}(1+h_{xx}\left((1+h_{xx})^{\frac{\gamma -1}{3-\gamma}}-1\right).
\end{equation}
\pagebreak 

From Eq.~\eqref{Spm}, we also have 
\begin{eqnarray}\label{Simple}
& S_+=\frac{(\gamma +1)}{4c_{s(0)^2}}\left\{\delta J_+\frac{\partial^2\delta J_+}{\partial x\partial t}- c_{s0}\delta J_+\frac{\partial^2\delta J_+}{\partial x^2}\right\}\nonumber\\
&+\frac{(\gamma +1)}{4c_{s(0)^2}}\left\{\left(\frac{\partial \delta J_+}{\partial t}\right)\left(\frac{\partial \delta J_+}{\partial x}\right)- c_{s0}\left(\frac{\partial \delta J_+}{\partial x}\right)^2\right\},\\
& S_-=0.
\end{eqnarray}
Limiting ourselves to second order perturbations, 
\begin{multline}\label{S}
\hspace*{-1em}S_{\mu\nu}=
\frac{(\gamma +1)}{2c_{s0}^2}((2-\gamma)h_{tx}+2c_{s0}h_{xx})(-c_{s0}\partial_x^2h_{\mu\nu}+\partial_x\partial_t h_{\mu\nu})\\
+\frac{(\gamma +1)}{2c_{s0}^2}(\partial_xh_{\mu\nu})(\partial_t-c_{s0}\partial_x)\left((2-\gamma)h_{tx}+2c_{s0}h_{xx}\right).
\end{multline}
However, the above expression can also be written in terms of $h_{xx}$ or $h_{xt}$, or a different linear combination of $h_{xx}$ and $h_{xt}$: In the case of simple waves, $h_{xx}$ and $h_{xt}$ are not independent due to relation \eqref{hxthxx}.

For a non-simple wave, we have to consider both $\delta J_+$ and $\delta J_-$ which are related to $h_{tx}$ and $h_{xx}$ by Eq.~\eqref{jh}. Therefore, we only look at the source terms for the 
$h_{tx}$ and $h_{xx}$ components, Note $h_{tt}$ is not an independent quantity, but is related to $h_{xt}$ and $h_{xx}$ via Eq.~\eqref{nhtt}. Using Eq.~\eqref{bf}, the source terms evaluated for $h_{xx}$ and $h_{tx}$ are, up to second power in $h_{\mu\nu}$, 
\begin{widetext}
\begin{multline}
 S_{xx}=\frac{r}{\Delta J_0}\left[\frac{(5-3\gamma)\Delta J_0}{4c_{s0}^2r}\frac{\partial^2}{\partial t\partial x}(h_{tx}h_{xx})-\frac{1}{c_{s0}}\frac{\partial^2}{\partial x^2}h_{tx}^2\right]
+\frac{r}{\Delta J_0} \left[-\frac{(\gamma -1)}{2c_{s0}}\left(\frac{\Delta J_0}{2r}\right)^2\frac{\partial^2}{\partial x^2}h_{xx}^2\right]\\
 +\frac{r(r-1)}{\Delta J_0^2}\left[\left(\frac{1}{c_{s0}}\frac{\partial h_{tx}}{\partial t}+\frac{\Delta J_0}{2r}\frac{\partial h_{xx}}{\partial x}\right)^2-\left(\frac{1}{c_{s0}}\frac{\partial h_{tx}}{\partial x}+\frac{\Delta J_0}{2rc_{s0}}\frac{\partial h_{xx}}{\partial t}\right)^2\right],
\end{multline}
\begin{multline}
 S_{tx}=S_{xt}=-\frac{1}{2}\left[\frac{1}{c_{s0}^2}\frac{\partial^2}{\partial t\partial x}h_{tx}^2+\left(\frac{\Delta J_0}{2r}\right)^2\frac{(\gamma -1)}{2c_{s0}^2}\frac{\partial^2}{\partial t\partial x} h_{xx}^2\right]
 -\left[\frac{1}{r}\frac{\partial^2}{\partial x ^2}(h_{tx}h_{xx})+\frac{\partial}{\partial x}(h_{tx}\partial_xh_{xx})\right],
\end{multline}
\end{widetext}
where we defined $r=\frac{3-\gamma}{\gamma -1}$.
\section{Calculating the mass flux}\label{massflux}
We here compute the time-averaged 
mass flux through $x=\xi_i$, at $\xi_i$, the boundary between source region and region of observation, $v=0$ initially. 
The mass flux in a one-dimensional flow is $j$ is given by
\begin{eqnarray}
&j=\rho v\nonumber\\
& =\rho_{0} v+\delta \rho v\nonumber\\
& \Rightarrow \left< j \right>=\rho_{0}\left< v \right>+\left< \delta\rho v \right> .
\end{eqnarray}
Taking into account up to second order in $v$ terms, and using Eq.~\eqref{rhov},
\begin{equation}\label{<j>S}
\left< j \right>\simeq\frac{\rho_{0}}{c_{s0}}(c_{s0}\left< v \right>+\left<  v^2 \right>).
\end{equation}
Here, $\left< v \right>$ at $x=\xi_i$  is given by
\begin{equation}\label{<v>S}
\left< v \right>=\frac{kc_{s0}}{2\pi}\int^{\frac{kc_{s0}}{2\pi}}_0 vdt=\frac{kc_{s0}}{2\pi}\int^{\xi_i-\lambda}_{\xi_i} f(\xi)\frac{1}{\partial_t\xi}d\xi,
\end{equation}
where $\partial_t\xi$, the partial derivative of $\xi$ with respect to $t$ is evaluated at $x=\xi_i$.

Using Eq.~\eqref{chh}, $\partial_t\xi$ at $x=\xi_i$ is given by
\begin{equation}
\partial_t\xi=-\left(\frac{c_{s0}+\frac{\gamma +1}{2} f(\xi)}{1+\frac{\gamma +1}{2} f'(\xi)\frac{\xi_i-\xi}{c_{s0}+
\frac{\gamma +1}{2}
 f(\xi)}}\right).
\end{equation}
We have the initial cosine profile, $f(\xi)=A\cos k\xi$.
Considering again terms up to second order, now in $f(\xi)$, we evaluate the integral in Eq.~\eqref{<v>S}, and after some further manipulations, we find from Eq.~\eqref{<j>S} the final result 
for the averaged mass flux
\begin{equation}
\left< j \right >=\rho_{0}\frac{3-\gamma}{c_{s0}}\frac{A^2}{8},
\end{equation}
which is Eq.~\eqref{A^2} in the main text. 
\end{appendix}

\bibliography{cqg_v16}

\end{document}